\documentclass[conference]{IEEEtran}
\IEEEoverridecommandlockouts
\usepackage{cite}
\usepackage{amsmath,amssymb,amsfonts}
\usepackage{algorithmic}
\usepackage{graphicx}
\usepackage{textcomp}
\usepackage{verbatim}
\usepackage{tabularx} 
\usepackage{pifont}
\usepackage[colorlinks=true, allcolors=blue]{hyperref}
\usepackage{xcolor}
\def\BibTeX{{\rm B\kern-.05em{\sc i\kern-.025em b}\kern-.08em
    T\kern-.1667em\lower.7ex\hbox{E}\kern-.125emX}}
\begin{document}

\title{ Why No Consensus on Consensus? A Deep Dive into Blockchain Consensus Protocols\\}

\author{\IEEEauthorblockN{Mohammad Pishdar}
\IEEEauthorblockA{\textit{1-VAISR Research Group, Tehran, Iran} \\
\textit{2- IDEAS Group, DistriNet, KU Leuven, Leuven, Belgium}\\mohammad.pishdar@kuleuven.be}
\and
\IEEEauthorblockN{Jawad Manzoor}
\IEEEauthorblockA{\textit{Department of Computer Science} \\
\textit{University of Galway}\\
Galway, Ireland \\
jawad.manzoor@universityofgalway.ie}
}

\maketitle

\begin{abstract}
Blockchain technology has revolutionized the digital landscape, driving innovations across industries through its decentralized and transparent infrastructure. These networks are primarily categorized as {\em public} or {\em private}, based on user access permissions. Public blockchains are open to all and fully decentralized while private blockchains have restricted access to authorized participants only and they are usually centralized or partially decentralized. Consensus protocols are at the heart of blockchain networks, playing a pivotal role in maintaining security, ensuring consistency, and achieving agreement among distributed nodes.
This paper provides a critical and unified analysis, including detailed workflows, that addresses the limitations of recent literature.
Furthermore, this research investigates the strengths and limitations of each protocol, shedding light on their suitability for various applications, including financial transactions, supply chain management, healthcare, and beyond. A critical analysis of ongoing challenges, such as security vulnerabilities, scalability bottlenecks, and energy consumption is provided. Finally, the paper identifies key research gaps in the field, offering insights into potential areas for future work and emerging trends aimed at addressing these issues. This comprehensive analysis serves as a valuable resource for researchers, practitioners, and organizations seeking to understand the role of consensus protocols in shaping the future of blockchain technology.

\end{abstract}


\section{Introduction}

Blockchain technology has evolved from its cryptocurrency origins to provide transformative solutions across industries like finance, supply chain, healthcare, and governance \cite{zhang2024novel,zeadally2019blockchain,arora2022blockchain}. Its core attributes are decentralization, immutability, and transparency, which enable trustless systems for secure data exchange, automated contracts (smart contracts), and auditable record-keeping \cite{chen2018survey,singh2020blockchain}. Global blockchain market capitalization now exceeds \$2.35 trillion \cite{mishra2021evolution,noauthor_undated-bv}, underscoring its economic significance.

Functionally, blockchain operates as a distributed ledger that exchanges data across a network of interconnected nodes. At the heart of blockchain functionality lies the consensus protocol, a mechanism ensuring agreement among distributed nodes on transaction validity and ledger state \cite{pahlajani2019survey,rebello2024survey}. These protocols directly govern critical performance dimensions: security, scalability, throughput, and energy efficiency. Public blockchains (e.g., Bitcoin, Ethereum) leverage permissionless consensus, e.g., such as Proof of Work (PoW), Proof of Stake (PoS), but face scalability-energy trade-offs \cite{strehle2020public}. Private blockchains, prioritizing speed and control, adopt efficient alternatives like Practical Byzantine Fault Tolerance (PBFT) \cite{kwak2019study}.
Despite progress, challenges persist in balancing decentralization with performance, mitigating attacks, and reducing resource consumption \cite{rebello2022security,bouraga2021taxonomy}.

While consensus protocols are well-studied individually, comparative analyses across both public and private contexts, focusing on operational workflows, security-scalability trade-offs, and application-specific suitability, remain underexplored. This paper addresses this gap with key contributions, including a comparative analysis of workflows of major public and private blockchain consensus protocols, an assessment of each protocol’s security, scalability, energy footprint, and suitability for key applications, and identification of emerging trends and future research avenues to overcome current limitations.

The structure of the paper is organized as follows. Section 2 provides an overview of essential concepts in blockchain networks. Section 3 reviews the related work. Sections 4 and 5 present the workflows of consensus protocols in public and private blockchains, respectively. Section 6 provides a comparative analysis of these protocols. Section 7 provides
a critical discussion and Section 8 identifies gaps in current research and possible future directions, followed by concluding remarks.

\begin{figure}
    \centering
    \includegraphics[width=1\linewidth]{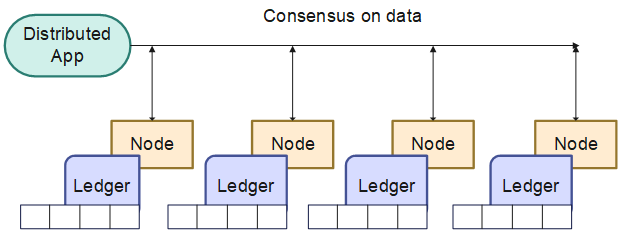}
    \caption{ General structure of a distributed ledger}
    \label{fig:DLT}
\end{figure}

\section{Background}
This section establishes foundational concepts essential for understanding consensus protocol analysis. We define core blockchain components, emphasizing their interplay in decentralized systems.

\begin{itemize}
\item \textbf{Distributed Ledger Technology (DLT):} A decentralized database architecture where identical copies of data are maintained across networked nodes. Unlike replicated databases, DLT synchronizes state through consensus algorithms without central authorities, ensuring tamper-resistance and transparency \cite{sunyaev2020distributed,buterin2014next}. Blockchain represents a specialized DLT implementation with unique structural properties. A sample structure of a distributed ledger is illustrated in Figure \ref{fig:DLT}.

\item \textbf{Blockchain Architecture:} A chained sequence of cryptographically linked blocks. Each block contains: (1) transaction batch, (2) timestamp, (3) nonce, and (4) hash pointers linking to previous/next blocks. This creates an append-only, immutable structure where each block is validated and added to the chain through a consensus mechanism, ensuring network-wide agreement \cite{singh2020blockchain,mourtzis2023blockchain}.

\item \textbf{Transactions and Blocks:} Transactions are digitally signed data units (e.g., asset transfers, contract calls). Transactions are aggregated into blocks of a certain size. Block size and generation intervals are protocol-specific parameters \cite{mourtzis2023blockchain}. 

\item \textbf{Cryptographic Hashing:} Cryptographic hashing is core to blockchain integrity. Algorithms like SHA-256 generate fixed-length outputs (hashes) from variable inputs. Each block contains:  
\( \text{Hash}_{\text{current}} = \mathcal{H}(\text{Hash}_{\text{prev}} \parallel \text{Transactions} \parallel \text{Nonce}) \)  
where \(\mathcal{H}\) is the hashing function. Any data change invalidates all subsequent hashes \cite{zheng2018blockchain,anceaume2020finality,xiao2020survey}.

\item \textbf{Transaction Finality:} It refers to the point at which a transaction is permanently recorded in the blockchain and can no longer be altered or reversed. Approaches include:  
\\- \textit{Probabilistic} (PoW/PoS): Confidence increases with block depth  
\\- \textit{Absolute} (BFT-style): Instant irreversibility.

Finality models directly impact security-latency tradeoffs \cite{anceaume2020finality,xiao2020survey}.

\item \textbf{Consensus Protocols:} A transaction requires a collective consensus before it can be definitively registered in the blockchain-distributed system. Consensus protocols ensure agreement among distributed nodes regarding the validity of transactions and blocks. These protocols verify transaction details, such as the sender’s digital signature, transaction amount, and availability of resources. Once a majority of nodes approve a transaction, it is added to the blockchain. Popular consensus mechanisms include PoW, PoS, PBFT, among others. Each protocol varies in terms of performance, security, and scalability, making them suitable for different blockchain applications  \cite{xiao2020survey,bouraga2021taxonomy}.

\item \textbf{Smart Contracts:} 
Smart contracts are self-executing agreements with their terms directly written into code. They execute autonomously on-chain when predefined conditions are met. 
In a simple transaction, for instance, the buyer immediately becomes the owner of an item, such as digital art or real estate, after the payment is received, thanks to the smart contract. Smart contracts can be executed by the Ethereum Virtual Machine (EVM) or other blockchain-specific runtime environments. They are usually written in high-level programming languages like Chaincode (for Hyperledger Fabric) or Solidity (for Ethereum) \cite{alaba2024smart,sun2024gptscan,pishdar2024major}.

\item \textbf{Network Types:}  Blockchains can broadly be categorized into public and private types.
\\- \textit{Public Blockchains}: Public blockchains are permissionless decentralized networks with open participation (e.g., Bitcoin, Ethereum). They rely on economic incentives (token rewards) and prioritize transparency and decentralization, but often face challenges related to scalability and energy consumption \cite{strehle2020public,ferdous2021survey}.
\\- \textit{Private Blockchains}: Private blockchains are permissioned and restrict access to authorized participants and are often centralized or partially decentralized. They have identity-based access control and use efficient consensus (e.g., PBFT, Raft) for higher throughput and lower latency. Ideal for enterprise consortia requiring governance \cite{strehle2020public,pahlajani2019survey}.  
\end{itemize}

\begin{table*}[t]
\centering
\caption{Comparative Analysis of Consensus Protocol Surveys }
\label{tab:survey_comparison}
\renewcommand{\arraystretch}{1.2} 
\begin{tabularx}{\textwidth}{l|c|c|c|c|c}
\textbf{Study} & \textbf{Protocol Coverage} & \textbf{Workflow Diagrams} & \textbf{Public/Private Balance} & \textbf{Security Analysis} & \textbf{Application Guidance} \\
\hline
Xiao et al. (2020) & Moderate & \ding{55} & Partial & Basic & Limited \\
Bouraga et al. (2021) & Broad & \ding{55} & Partial & Moderate & Theoretical \\
Guru et al. (2023) & Moderate & \ding{55} & Partial & Focused & Minimal \\
\hline
\textbf{Shen et al. (2025)} & Broad & \ding{55} & Partial & High (Performance) & Limited \\
\textbf{Pineda et al. (2024)} & Broad & \ding{55} & Partial & Minimal (Energy Focus) & Theoretical \\
\textbf{Jain et al. (2025)} & Narrow (Scalability) & \ding{55} & Partial & Basic & Limited \\
\hline
\textbf{Our Work} & \textbf{Comprehensive} & \textbf{\ding{51}} & \textbf{\ding{51}} & \textbf{Systematic} & \textbf{Multi-domain} \\
\end{tabularx}
\end{table*}

\section{Related Work}
This section critically examines prior research on blockchain consensus protocols, categorizing studies by methodological approach and highlighting limitations that motivate our work. We structure our analysis into four thematic clusters: general surveys, application-specific reviews, classification frameworks, and security analyses.

\subsection{General Surveys and Overviews}
While existing surveys do indeed provide foundational insights, there are also noticeable gaps in depth and methodological rigor, and many display a persistent tendency toward descriptive enumeration rather than critical, comparative analysis. Foundational works such as Xiao et al. \cite{xiao2020survey}, Sankar et al. \cite{sankar2017survey}, and Wahab et al. \cite{wahab2018survey} provide useful description at the algorithmic level but fall short with respect to exhaustive operational mechanisms, protocolagnostic visualizations, and a deep evaluation of the core trade-offs and implementation challenges. Studies of narrower scope also suffer from severe limitations: Kaur et al. \cite{kaur2021research} evaluate only six protocols while Tosh et al. \cite{tosh2017consensus} restrict their coverage to PoW and PoS. Ismail et al. \cite{ismail2019review} introduce a compute-centric taxonomy but omit critical dimensions such as finality models and decentralization trade-offs. Altogether, these seminal contributions lacked standardized evaluation frameworks and a balanced coverage of public and private consensus mechanisms.

Despite the proliferation of literature in 2024 and 2025, a comprehensive, unified assessment is still lacking, since recent surveys often show over-specialization and fragmentation in their analytic focus. Comprehensive reviews, such as Shen et al. (2025) \cite{shen2025blockchain} and Al-awamy et al. (2025) \cite{al2025hybrid}, rely heavily on technical performance frameworks and hybrid protocols but often fail to connect these metrics to practical domain-specific selection guidance. Furthermore, several recent systematic studies show highly restrictive scope: for example, Pineda et al. (2024) \cite{pineda2024sustainable} focuses on sustainability and energy consumption while Jain et al. (2025) \cite{jain2025survey} focuses strongly on scalability and Layer-2 solutions, underrepresenting the security and decentralization properties of the underlying Layer-1 protocols. Other approaches remain abstract, with the bibliometric analysis of Ahn et al. (2024) \cite{ahn2024blockchain} providing statistical trends but without practical operational insights, and performance-centric analyses like Chan et al. (2025) \cite{chan2025performance} remain narrow in their protocol scope. Most importantly, even regulatory-centric primers like Bains (2025) \cite{bains2025blockchain} lack the necessary technical depth. This persistent fragmentation and lack of a unified actionable guide justify the need for our work.

\subsection{Classification and Framework-Based Studies}
Taxonomic approaches provide good conceptual organization but insufficient practical guidance. Bouraga et al. \cite{bouraga2021taxonomy} develop a robust security-performance classification but omit workflow visualizations and empirical validation. Bano et al. \cite{bano2019sok} categorize protocols into three archetypes yet fail to translate their framework into actionable design principles. Zhang et al. \cite{zhang2020analysis} innovatively classify by finality (probabilistic vs. absolute) but overlook hybrid mechanisms emerging in modern systems (e.g., Ethereum's hybrid PoS/PBFT). Moreover, the recent literature is fragmented toward highly specialized classification; for example, the SoK analysis of Raikwar et al. (2024) \cite{raikwar2024sok} focuses exclusively on DAG-based consensus protocols-a useful study in its niche that is fundamentally lacking in scope to systematize the broader family of chain-based consensus mechanisms.

While these studies are useful in systematizing protocol attributes, they collectively share three critical limitations: (1) minimal cross-protocol benchmarking, (2) abstract treatment of operational workflows, and (3) insufficient connection to real-world deployment constraints.

\subsection{Application-Specific Consensus Analyses}
Domain-focused studies yield specific insights but lack generalizability. Tomic et al. \cite{tomic2021review} thoroughly evaluate private blockchain consensus yet provide no comparative visualizations or scalability metrics applicable beyond enterprise contexts. Salimitari et al. \cite{salimitari2018survey} effectively map IoT requirements to consensus properties and explore the compatibility of consensus protocols with IoT networks, aiming to identify the most suitable methodology. However, it is limited in scope to a single domain, overlooking general-purpose blockchain applications. Bashar et al. \cite{bashar2019contextualizing} offer valuable market-driven analysis of top cryptocurrencies but exclude permissioned networks and emerging protocols.

\subsection{Security and Attack-Centric Reviews}
Vulnerability-focused research addresses critical gaps but remains fragmented. Guru et al. \cite{guru2023survey} catalog attack vectors against consensus layers but do not cover protocol mechanics or mitigation strategies. Cachin et al. \cite{cachin2017blockchain} provide rigorous Byzantine fault tolerance analysis for private networks while neglecting public blockchain threat models. Neither study systematically evaluates how protocol designs inherently enable or resist attacks across network types.

\subsection{Research Gaps and Our Contributions}
The existing literature, even among the most recent 2024 and 2025 surveys, exhibits five persistent limitations:

\begin{itemize}
    \item \textbf{Scope Imbalance:} Underrepresentation of private blockchain protocols vs. public counterparts, a balance often skewed in favor of public blockchains in general surveys.
    \item \textbf{Methodological Gaps:} Insufficient visual workflow documentation and comparative benchmarking, a core deficiency noted even in performance-centric analyses like Chan et al. (2025) \cite{chan2025performance}.
    \item \textbf{Evaluation Fragmentation:} Lack of unified security-scalability-decentralization assessment criteria, as evidenced by highly specialized recent reviews that focus on a single dimension, such as the sustainability focus of Pineda et al. (2024)\cite{pineda2024sustainable}.
    \item \textbf{Evolutionary Blindspots:} Limited balanced coverage of hybrid/novel protocols (e.g., Proof of Elapsed Time (PoET), YAC) alongside foundational Layer-1 protocols, leading to fragmentation (e.g., Raikwar et al. 2024 \cite{raikwar2024sok}.
    \item \textbf{Practical Disconnect:} Minimal guidance on protocol selection for domain-specific requirements, rendering otherwise comprehensive reviews (e.g., Shen et al. 2025 \cite{shen2025blockchain} insufficiently actionable for system architects.
\end{itemize}
Our study directly addresses these persistent gaps through the following contributions, providing the unified, critical, and actionable framework missing from the current state-of-the-art literature:

\begin{itemize}
    \item \textbf{Comprehensive Protocol Coverage:} Analyzing 9 consensus mechanisms (PoW, PoS, Delegated Proof-of-Stake (DPoS), PBFT, Raft, YAC, PoET, Kafka, Paxos) across public and private blockchains, providing the necessary balanced scope.
    \item \textbf{Operational Transparency:} Providing standardized workflow diagrams and algorithmic breakdowns for each protocol, overcoming the abstract treatment in existing taxonomic works.
    \item \textbf{Unified Evaluation Framework:} Assessing protocols using consistent metrics: security, scalability limits, energy efficiency, and decentralization trade-offs, explicitly bridging the evaluation fragmentation present in specialized reviews.
    \item \textbf{Application-Specific Recommendations:} Mapping protocol strengths to use cases (DeFi, supply chain, healthcare etc.), thus providing the actionable guidance that resolves the Practical Disconnect.
    \item \textbf{Emerging Trend Synthesis:} Identifying research directions including cross-shard consensus, quantum-resistant designs, and incentive-alignment mechanisms.
\end{itemize}

\begin{figure*}
\centering
  \includegraphics[width=0.8\textwidth]{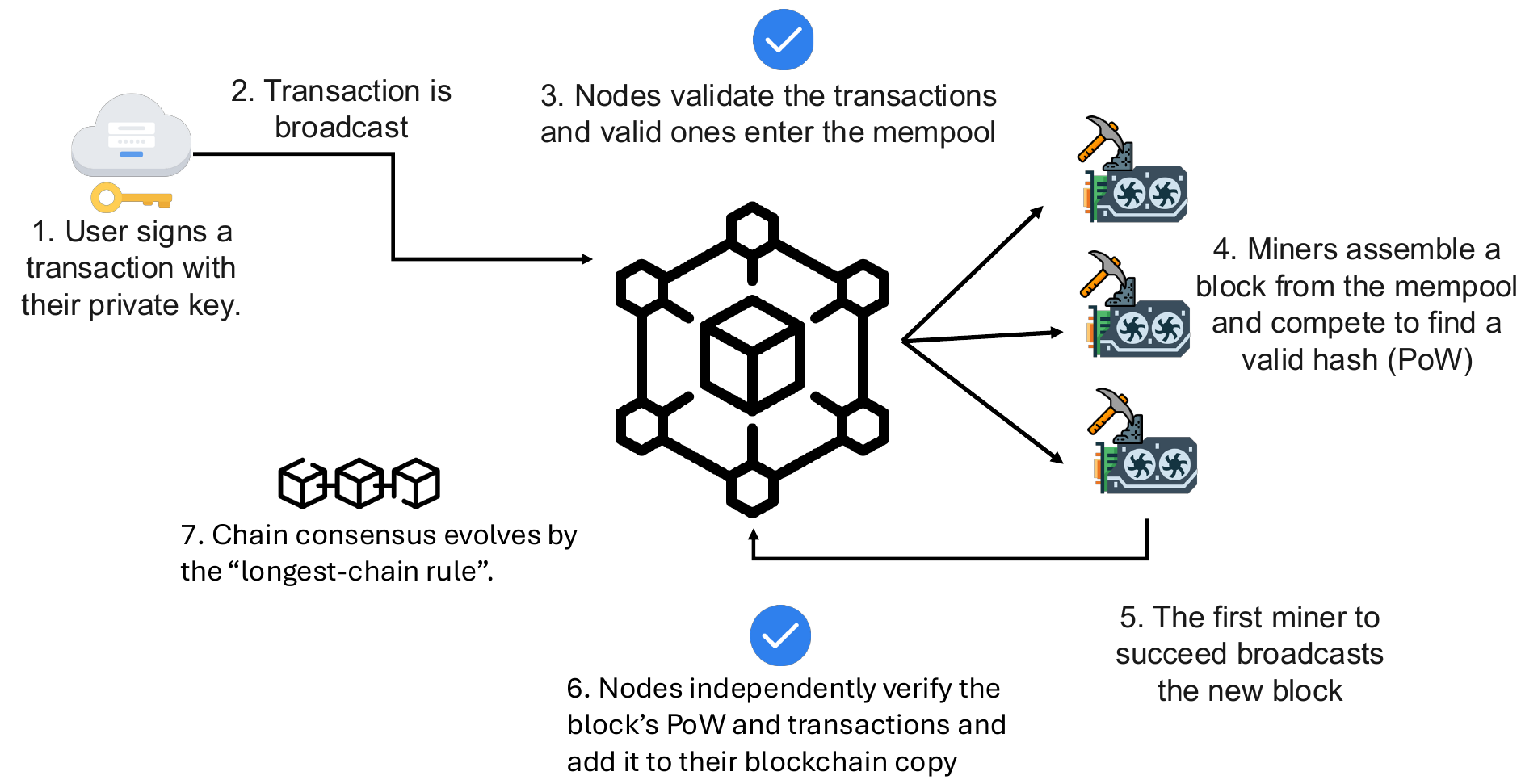}
  \caption{The workflow of a transaction in PoW consensus.}
  \label{fig:pow_transaction}
\end{figure*}

\begin{figure}
    \centering
    \includegraphics[width=1\linewidth]{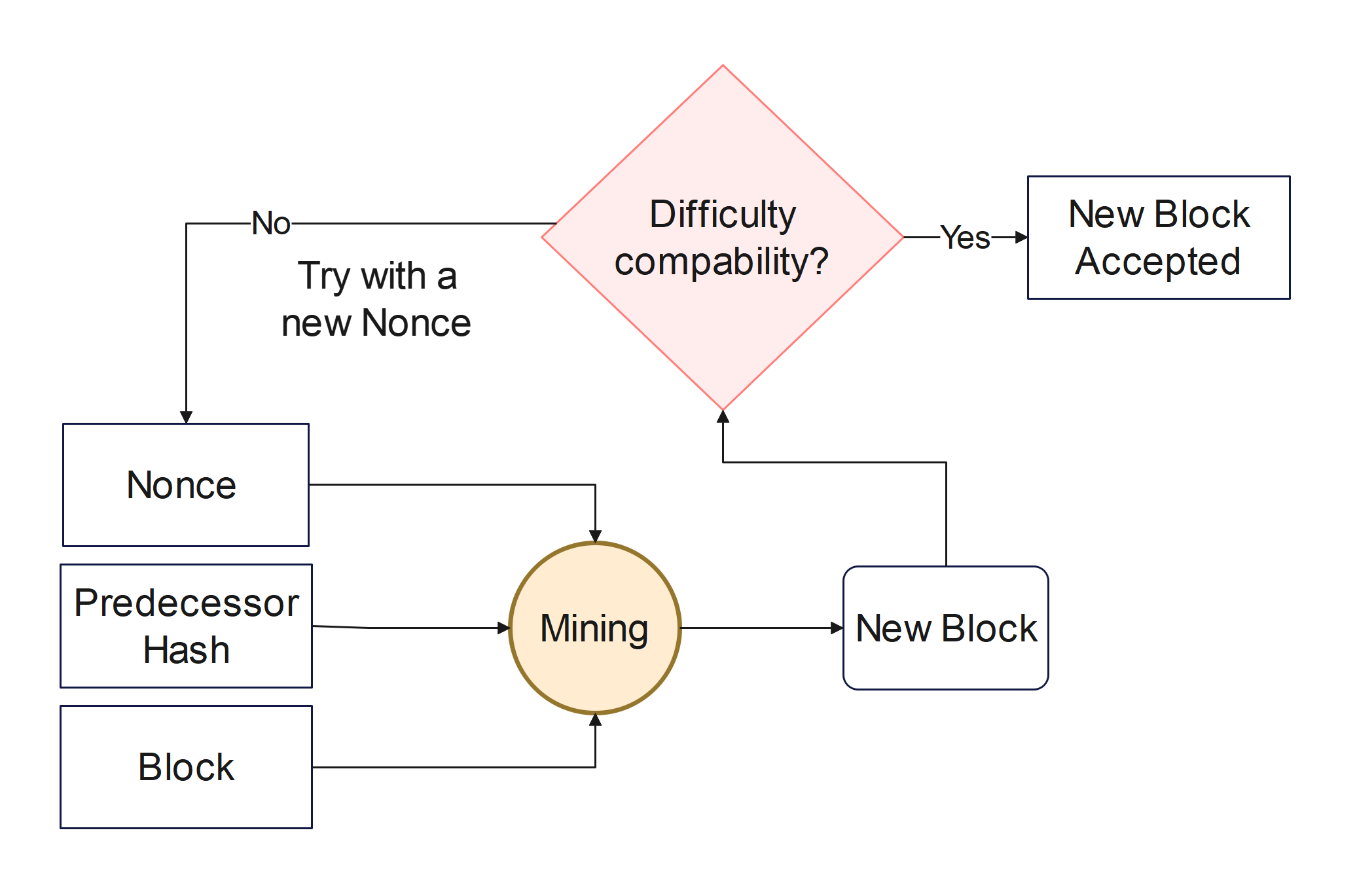}
    \caption{Block mining in PoW.
}
    \label{fig:pow}
\end{figure}

\section{\textbf{Consensus Protocol for public blockchains }
}

Public blockchains operate as open, permissionless networks where anyone can join, participate, and interact without requiring prior approval. Their architecture is designed around decentralization, transparency, and resistance to censorship, enabling participants across the world to transact and collaborate without relying on a central authority. Trust is established through decentralized mechanisms rather than pre-verified identities. Nodes maintain the blockchain by propagating transactions, verifying them, storing the full or partial ledger, and participating in the consensus process. The roles of nodes vary depending on the protocol: some may only relay data, while others, such as miners or validators, contribute computational or economic resources to secure the network.
Consensus is achieved through open participation mechanisms such as PoW, PoS, or other resource-based protocols. These mechanisms ensure that all participants agree on the state of the ledger, even in an environment with no trusted identities and potentially malicious actors. The process typically involves proposing blocks, validating them according to protocol rules, and finalizing them through network-wide agreement.
Once validated, blocks are added to the chain, forming an immutable, append-only ledger that is replicated across thousands of independent nodes. This global distribution provides strong guarantees of integrity, security, and fault tolerance, allowing public blockchains to function reliably without central oversight.
Some of the most widely used public blockchain consensus protocols are discussed next. 

\subsection{Proof of Work}

PoW, which was pioneered by Bitcoin \cite{noauthor_undated-ij,jin2017blockndn,nakamoto2008bitcoin,schinckus2021proof}, secures the network through a process of competitive computation. 
The detailed transaction workflow of PoW is illustrated in Figure \ref{fig:pow_transaction}. The process begins with a user initiating a transaction via their wallet. Following cryptographic signing, the transaction is broadcast to the network, where the nodes perform initial checks. Valid transactions are then propagated to the mempools of other nodes, where they await inclusion in a block. 
Miners construct a candidate block by selecting valid transactions from their mempool. They repeatedly attempt to solve a cryptographic puzzle by identifying a nonce $N$ such that the hash of the block header falls below a network-specified difficulty target, formally expressed as:
\[
\mathcal{H}(\text{PrevHash} \parallel \text{TxRoot} \parallel N) < \text{Target}
\]
where $\mathcal{H}$ denotes the cryptographic hash function, $\text{PrevHash}$ is the hash of the previous block, $\text{TxRoot}$ represents the Merkle root of all transactions in the block, and $\text{Target}$ encodes the current network difficulty.
This process, illustrated in Figure \ref{fig:pow}, is computationally intensive and energy-consuming but provides robustness. Once a miner finds a valid block, they broadcast it. The new block undergoes a verification process by the broader network. Finally, upon receiving sufficient confirmations, the transaction is considered finalized and is immutably appended to the distributed ledger.

\begin{figure*}
\centering
  \includegraphics[width=0.75\textwidth]{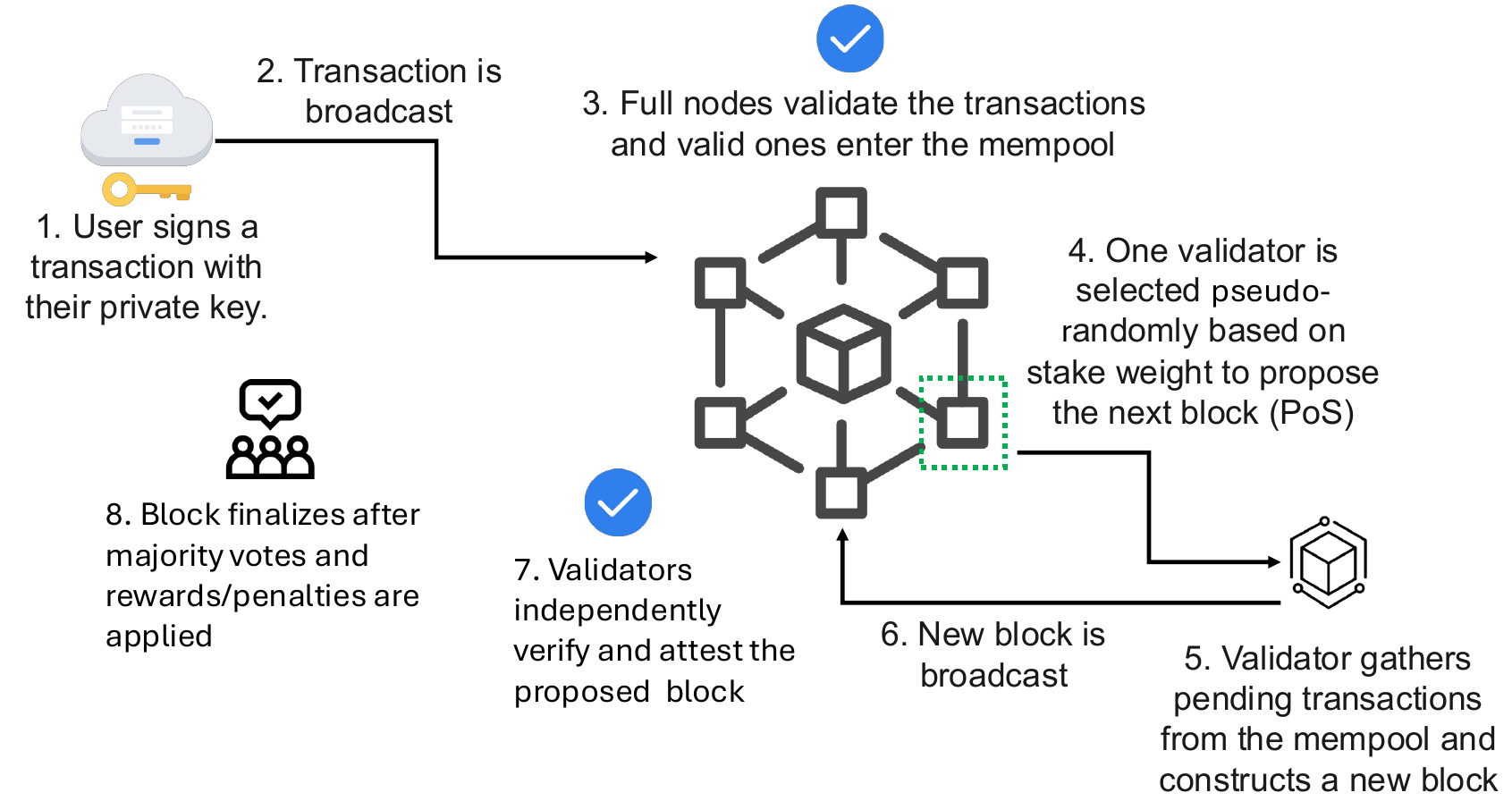}
  \caption{The workflow of a transaction in PoS consensus.}
  \label{fig:pos}
\end{figure*}

\begin{figure*}
\centering
  \includegraphics[width=0.75\textwidth]{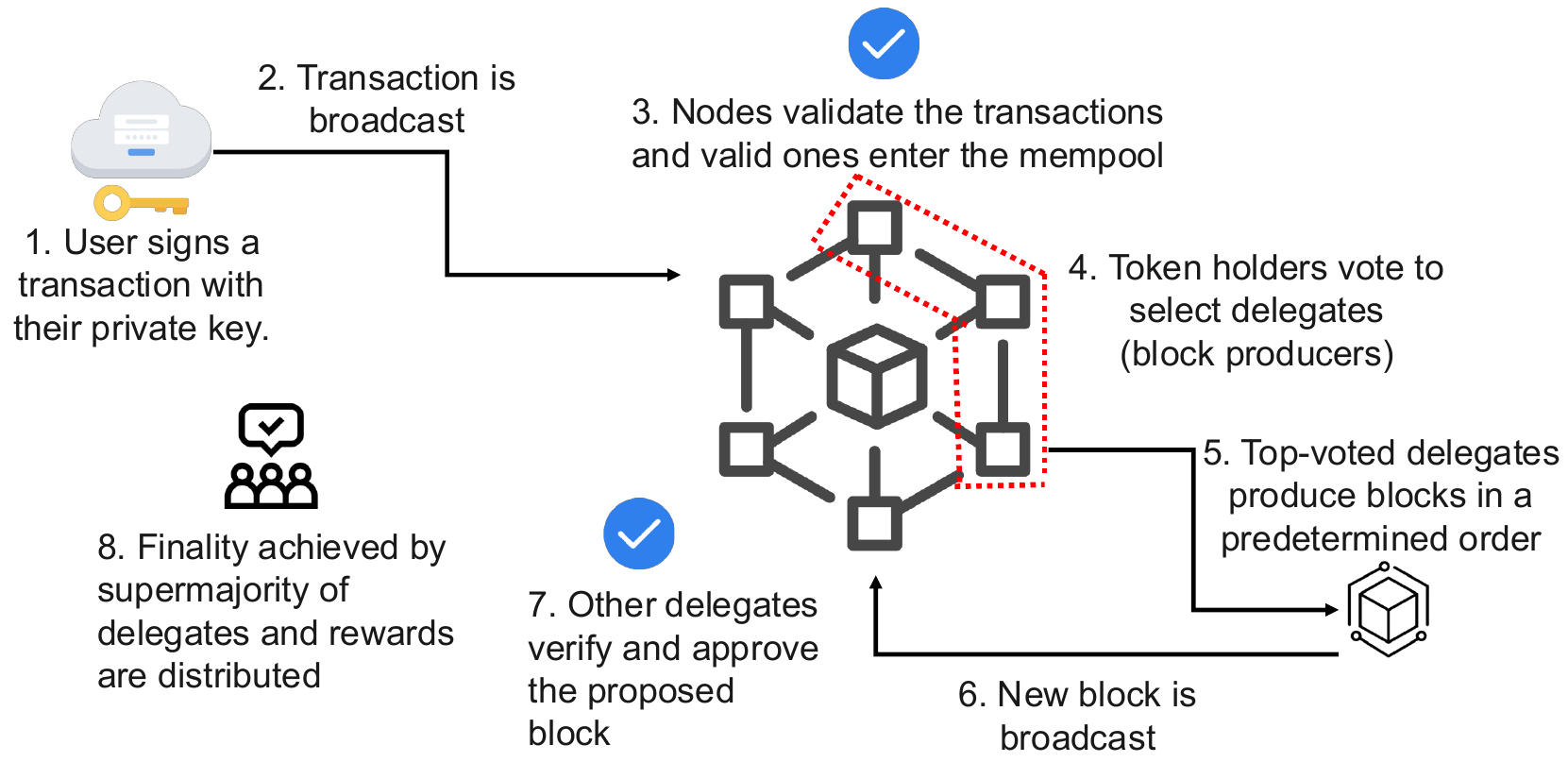}
  \caption{ The workflow of a transaction in DPoS consensus.}
  \label{fig:dpos}
\end{figure*}

The primary advantages of PoW are its well-established security and fully permissionless nature. However, these benefits come at the cost of high energy consumption and limited transaction throughput, presenting significant scalability challenges.

\subsection{Proof of Stake}
PoS emerged as an energy-efficient alternative to PoW \cite{saleh2021blockchain,bentov2014proof}. Instead of computational work, PoS protocols secure the network through economic stake. A validator's probability of being selected to propose the next block is proportional to the amount of cryptocurrency they have locked (staked) as collateral. This eliminates the need for energy-intensive mining, reducing energy consumption. The workflow is detailed in Figure \ref{fig:pos}, where the protocol selects a validator to write the next node based on the amount of staked assets and the waiting time of the nodes.
Other nodes then attest to the block's validity. To mitigate malicious behavior, PoS systems implement slashing mechanisms, where a validator’s staked funds may be partially or fully confiscated for misbehavior such as double-signing or proposing conflicting blocks. Furthermore, several modern PoS protocols incorporate finality gadgets (e.g., Casper FFG) to achieve absolute finality for certain checkpoints, making long-range attacks infeasible and providing stronger security guarantees than probabilistic finality alone.

Many modern PoS systems incorporate finality gadgets (e.g., Casper FFG) to provide absolute finality for certain checkpoints, preventing long-range attacks. While PoS offers superior energy efficiency and higher potential throughput, it introduces its own set of challenges, including potential wealth concentration which may lead to centralization.

\subsection{Delegated Proof of Stake}

DPoS further modifies the stake-based consensus model by introducing a representative democratic process \cite{yang2019delegated,hu2021improved}. In DPoS, token holders do not validate transactions themselves but instead vote to elect a limited number of delegates who are responsible for block production and network governance. These delegates typically take turns producing blocks in a round-robin fashion, leading to very fast block times and high throughput. As shown in Figure \ref{fig:dpos}, voters continuously monitor the performance of their chosen delegates and can reallocate their votes to others if they are dissatisfied, creating a system of real-time accountability. The primary advantage of DPoS is its performance; it enables high scalability and fast transaction finality. However, this comes at the cost of increased centralization, as block production is effectively entrusted to a small, elected cohort of nodes, making the network potentially vulnerable to cartel formation and reduced censorship resistance.

\section{\textbf{Consensus Protocol for Private blockchains }}

Private blockchains operate with permissioned access and typically aim for higher performance than public networks. Although they share basic concepts such as users and transactions, their architecture is shaped by a more structured governance model. Hyperledger Fabric is one of the most prominent examples, supported by major technology companies such as IBM, Intel, and SAP.\cite{fabric2020blockchain,bhuvana2020blockchain,noauthor_2024-xx,kaur2021research1}.

Access to network resources in private blockchains is managed through a permissions layer that specifies the roles and capabilities of users, nodes, and organizations. Identity and trust are established using cryptographic credentials, often issued by a certificate authority or a similar trusted identity management system that authenticates participants and supports accountability across the consortium.

Smart contracts in private blockchains serve as business logic components that automate workflows, enforce policies, and coordinate activities between members. Their design typically emphasizes efficiency, privacy, and adherence to governance rules. Access to contract functions can be restricted to specific roles or organizations, depending on the needs of the consortium.

Network nodes maintain the distributed ledger, validate transactions, and participate in the consensus process. The exact roles of nodes vary across platforms, but they generally include storing the ledger, executing or verifying smart contract logic, and contributing to block creation or transaction ordering. Consensus mechanisms used in private blockchains ensure that all organizations agree on the sequence and validity of transactions.

Once transactions have been validated and ordered according to the chosen consensus protocol, they are packaged into blocks and appended to the ledger maintained by each node. This process ensures a consistent, shared state across the consortium and completes the transaction lifecycle.

Next, we provide a detailed analysis of the prominent consensus protocols employed within this architecture.

\begin{figure*}
\centering
  \includegraphics[width=0.8\textwidth]{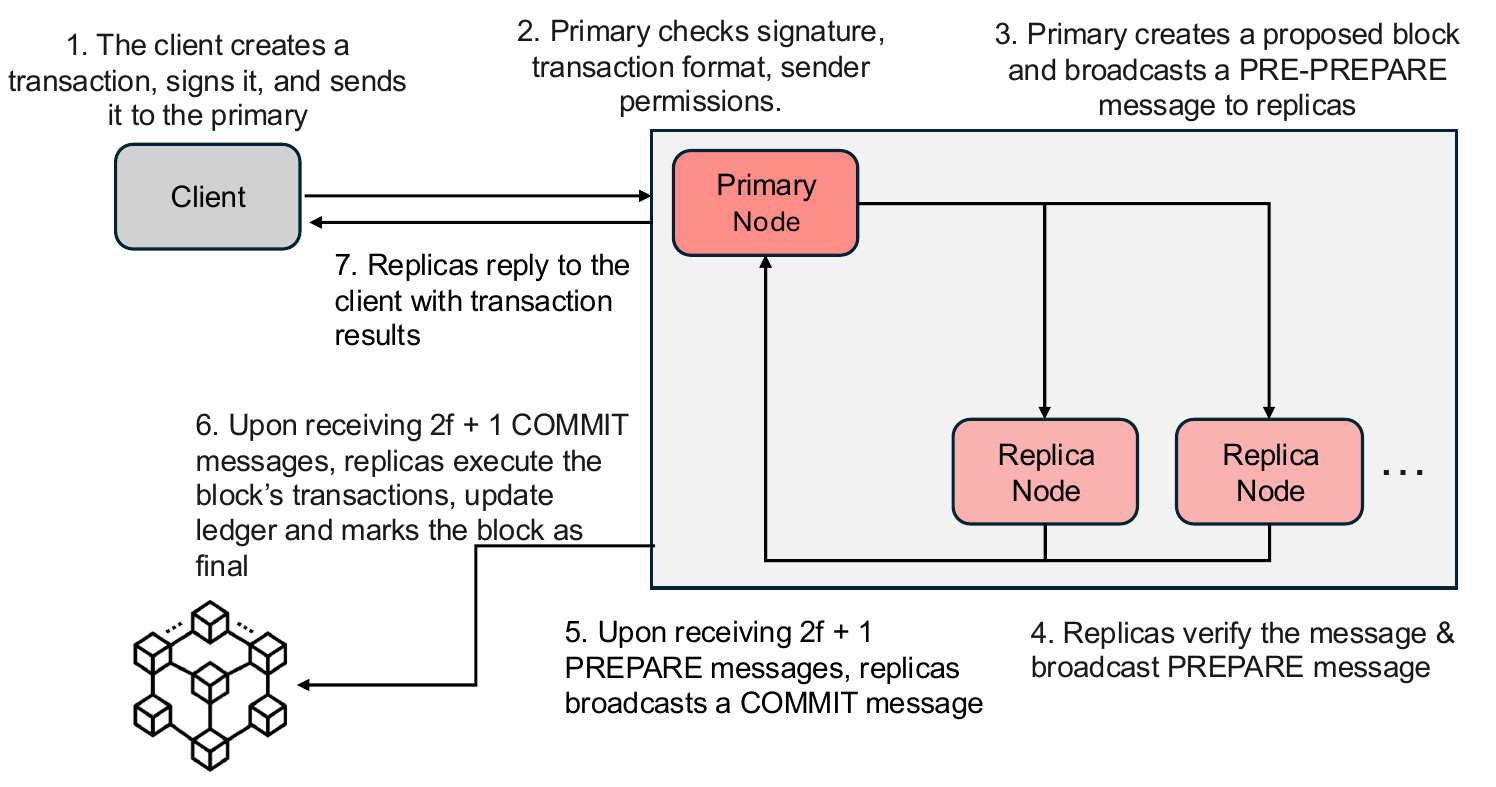}
  \caption{  Consensus workflow of a transaction in the PBFT protocol}
  \label{fig:pbft-workflow}
\end{figure*}

\subsection{Practical Byzantine Fault Tolerance}

PBFT is a classical consensus algorithm designed to efficiently tolerate Byzantine faults in a permissioned, crash-fault-tolerance-first setting \cite{abraham2017revisiting,gao2019t}. It operates in a series of phases orchestrated by a designated leader node, which is selected in a round-robin fashion. As illustrated in Figure \ref{fig:pbft-workflow}, the protocol proceeds as follows:

Pre-prepare: The leader assigns a sequence number to a client request and broadcasts a pre-prepare message to all backup nodes.

Prepare: Each node verifies the message and, if valid, broadcasts a prepare message. A node enters the prepared state upon receiving $2F + 1$ valid prepare messages, ensuring a sufficient number of nodes agree on the order of the request.

Commit: Nodes then broadcast commit messages. Upon receiving $2F + 1$ valid commit messages, a node knows that a sufficient quorum has committed to the decision. It then executes the request, updates its local state, and sends a reply to the client.

This three-phase commit ensures both safety (all honest nodes execute the same requests in the same order) and liveness (the system continues to process requests as long as at least $2F + 1$ nodes are honest) without the computational overhead of PoW. Its primary limitation is the communication complexity of $O(n^2)$ messages per request, which hinders its scalability to very large networks.

\begin{figure*}
\centering
  \includegraphics[width=0.8\textwidth]{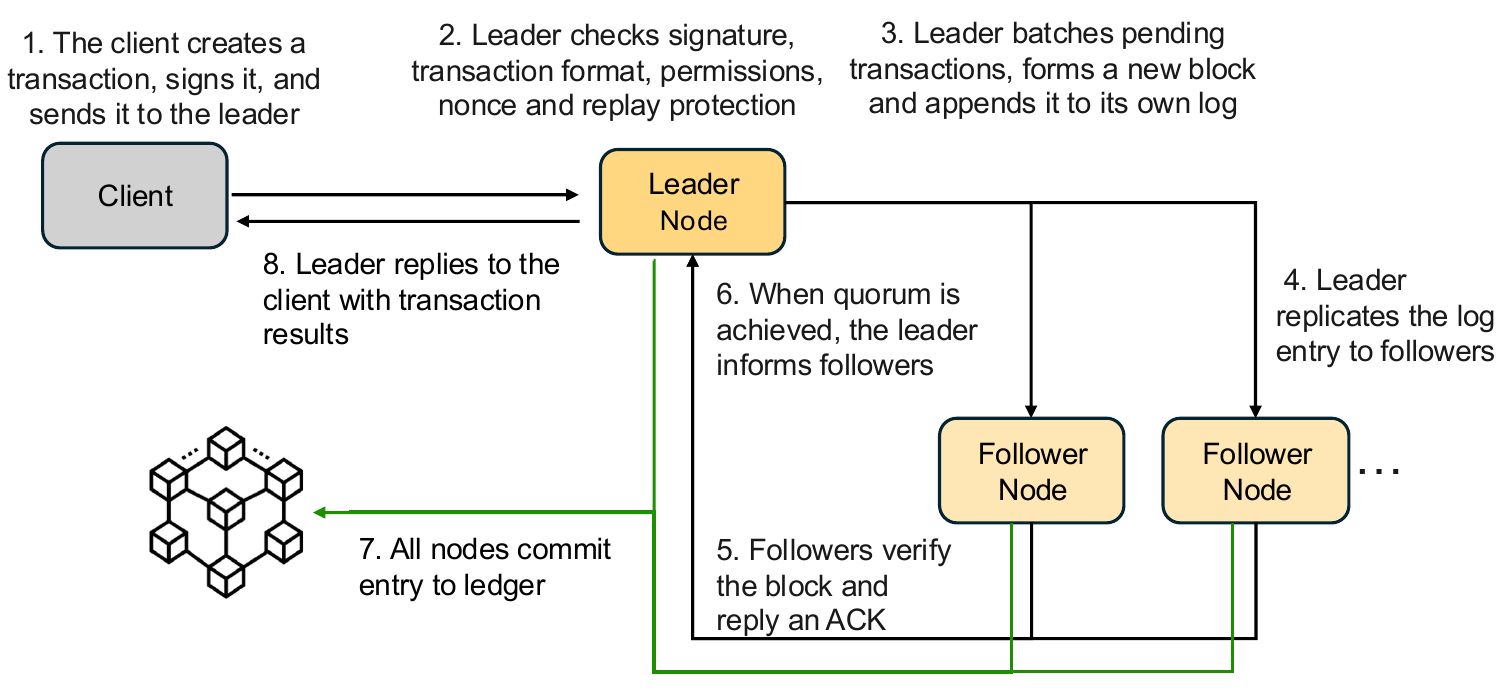}
  \caption{  Consensus workflow of a transaction in the Raft protocol}
  \label{fig:raft-workflow}
\end{figure*}

\subsection{Raft}

Raft is a consensus algorithm designed for understandability and ease of implementation. It is often favored in private blockchain networks for its simplicity and strong consistency guarantees in non-Byzantine environments \cite{hu2020raft,huang2019performance,ongaro2014search}. It organizes nodes into one of three states:

Leader: A single elected leader handles all client communication, replicates log entries to followers, and manages commit updates.

Follower: Passive nodes that respond to messages from the leader and candidate nodes.

Candidate: A transient state a node enters to initiate a leader election.

Raft operates through a leader-driven process. Leader election is triggered via timeouts; if a follower does not receive a heartbeat from the leader, it becomes a candidate and requests votes. A candidate wins the election and becomes the new leader if they receive votes from a majority of nodes. This strong leadership model makes Raft highly efficient and performant. However, its critical weakness is its non-Byzantine fault tolerance; it can only tolerate crash faults and is vulnerable to malicious leaders.

The workflow is illustrated in Figure \ref{fig:raft-workflow}. The protocol begins when the leader node receives and performs an initial validation of a client's transaction. It then propagates this transaction to all follower nodes for evaluation. The leader collects individual responses from these followers and verifies if a consensus majority has been reached. Upon achieving quorum, the leader commits the new block to its own local ledger and disseminates a confirmation block to all peers, instructing them to append it to their respective ledgers, thereby ensuring state consistency across the distributed network. Finally, the leader directly notifies the client of the transaction's outcome. This streamlined, leader-centric approach, which eliminates the computational overhead of competitive mining, is a primary reason for Raft's prevalence and ease of implementation in private blockchain environments.

\begin{figure*}
\centering
  \includegraphics[width=0.8\textwidth]{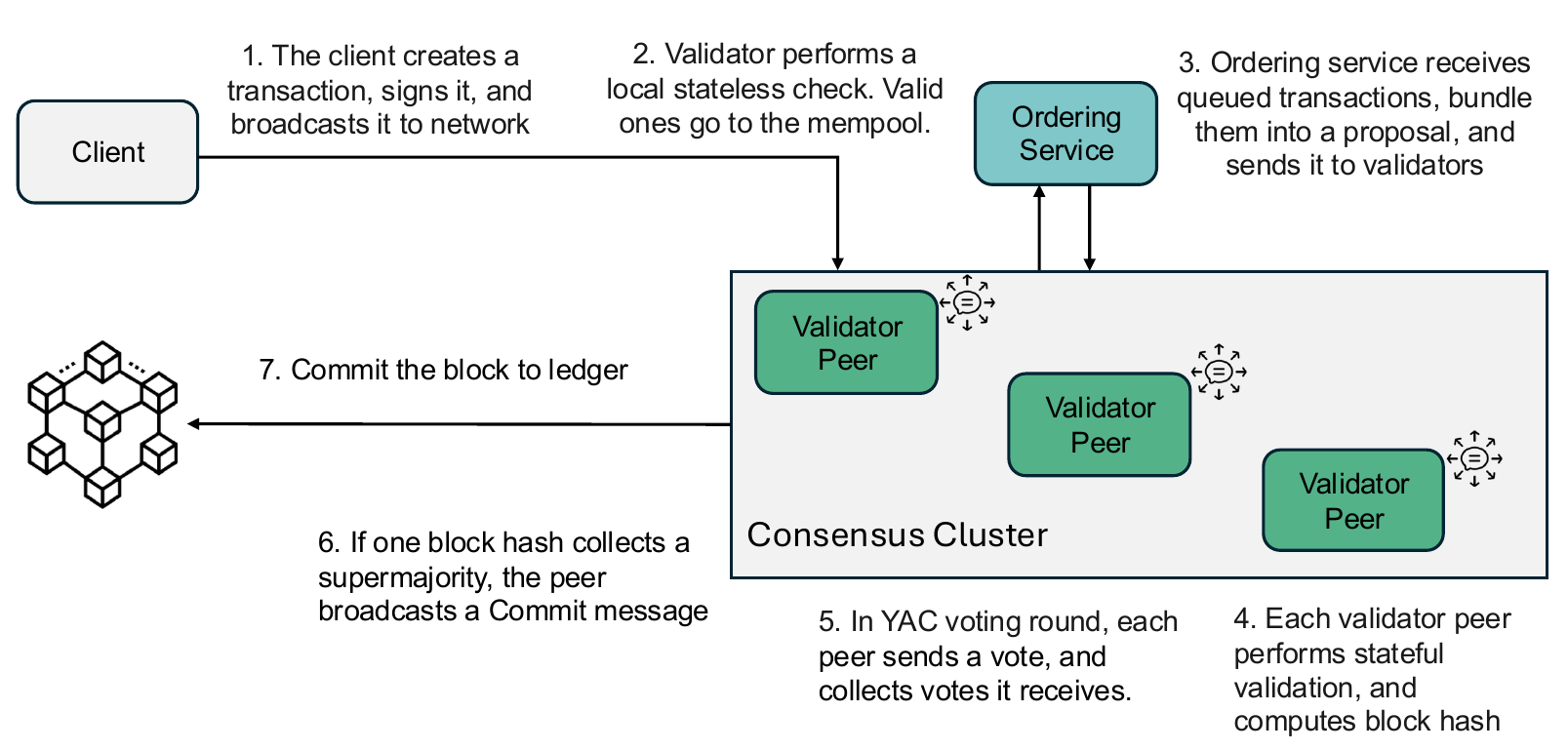}
  \caption{  Consensus workflow of a transaction in the YAC protocol}
  \label{fig:yac-workflow}
\end{figure*}

\begin{figure*}
\centering
  \includegraphics[width=0.9\textwidth]{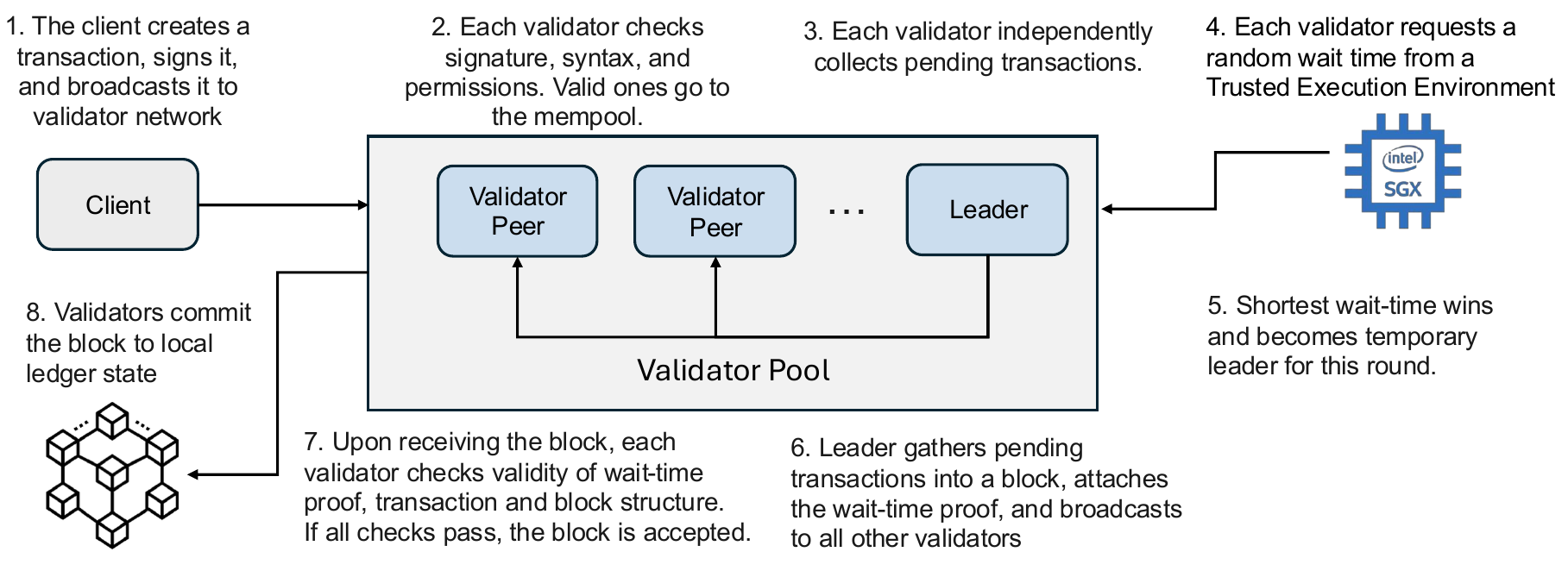}
  \caption{  Consensus workflow of a transaction in the PoET protocol}
  \label{fig:poet-workflow}
\end{figure*}

\subsection{YAC}

YAC is a consensus protocol that uses a deterministic, election-free ordering mechanism to reduce communication overhead \cite{muratov2018yac}. For each round, the order of nodes is calculated deterministically, for example using a function of the previous block’s hash and node identifiers. The transaction workflow of YAC, illustrated in Figure \ref{fig:yac-workflow}, involves the following steps:

A client submits a transaction to one or more peers. Each node validates the transaction and determines the deterministic voting order. Nodes then cast votes sequentially according to this order. As votes propagate through the network, any node that observes a supermajority of votes can commit the transaction block independently. Once committed, the block is propagated to all peers.

This design minimizes the message complexity typical of traditional BFT protocols by avoiding full all-to-all communication. However, the protocol’s security and performance critically depend on the robustness and fairness of the deterministic ordering function, as predictable or biased ordering could lead to centralization or manipulation.

\subsection{PoET}

PoET is a consensus algorithm that aims to achieve fair leader election through a trusted execution environment (TEE), specifically Intel's Software Guard Extensions (SGX) \cite{bouraga2021taxonomy,corso2019performance}. It operates on a principle similar to a verifiable lottery: each node requests a random wait time from its SGX enclave. The first node to complete its wait time is granted the right to propose a block. The node must then provide cryptographic proof generated by the SGX hardware that it indeed waited the required duration, which other nodes can efficiently verify. This process, illustrated in Figure \ref{fig:poet-workflow}, eliminates the need for competitive resource consumption, making it highly energy-efficient. The primary drawback of PoET is its heavy reliance on specialized hardware, creating a potential single point of failure and trust assumption (the integrity of Intel's SGX), which contradicts the trust-minimization goal of many decentralized systems.

\subsection{Kafka}

Kafka is a crash fault-tolerant (CFT) consensus protocol used for high-throughput ordering services, based on the Apache Kafka distributed streaming platform \cite{yang2022resource,kreps2011kafka,raptis2023survey}. It operates as a publish-subscribe system where:

\begin{itemize}
    \item Client applications are Producers that publish messages (transactions) to a Topic (an ordered log).
    \item Brokers (Kafka servers) store these topic partitions.
    \item The Ordering Service nodes act as Consumers, pulling ordered batches of messages from the Kafka topic.
\end{itemize}

\begin{figure*}
\centering
  \includegraphics[width=0.9\textwidth]{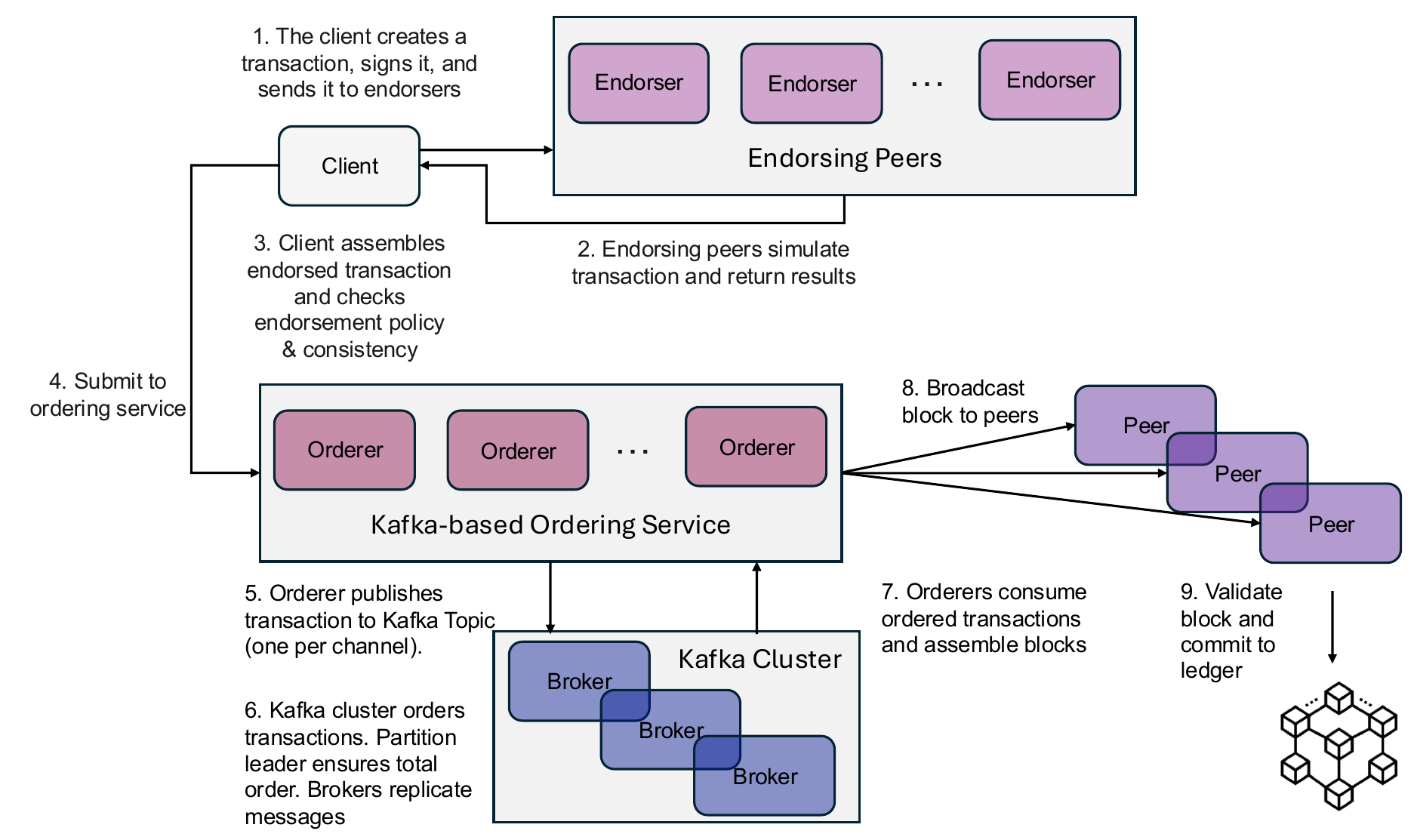}
  \caption{  Consensus workflow on a transaction in the KafKA protocol}
   \label{fig:kafka-workflow}
\end{figure*}

The Kafka consensus mechanism functions as a high-performance, distributed logging system to establish a total order of transactions. In this model shown in Figure \ref{fig:kafka-workflow}, a leader orderer node receives client transactions and forwards them to a leader broker, which replicates and persists the messages across a cluster of follower brokers, ensuring durability. A message is considered committed only after successful replication to a quorum of brokers, establishing consensus on its order and existence. Orderer nodes, acting as consumers, then periodically pull batches of these ordered messages from the broker leader using asynchronous requests that specify the offset of the last received message. The broker services these requests by retrieving the sequentially ordered messages from its segmented log files. An external Zookeeper service is integral to this architecture, as it manages cluster metadata, monitors broker health, and orchestrates leader election for both brokers and orderers to maintain availability and coordinate rebalancing in the event of node failures. Finally, the orderers collectively cut and assemble the totally ordered stream of messages into blocks, which are then disseminated to committing peers for ledger update. This design prioritizes extremely high throughput and low latency for ordering transactions, making it suitable for enterprise-grade private blockchain networks.  However, Kafka only provides CFT, relying on the assumption that all ordering nodes and brokers are honest. It is not suitable for adversarial environments where Byzantine behavior is a concern.

\subsection{Paxos}

\begin{figure*}
\centering
  \includegraphics[width=0.9\textwidth]{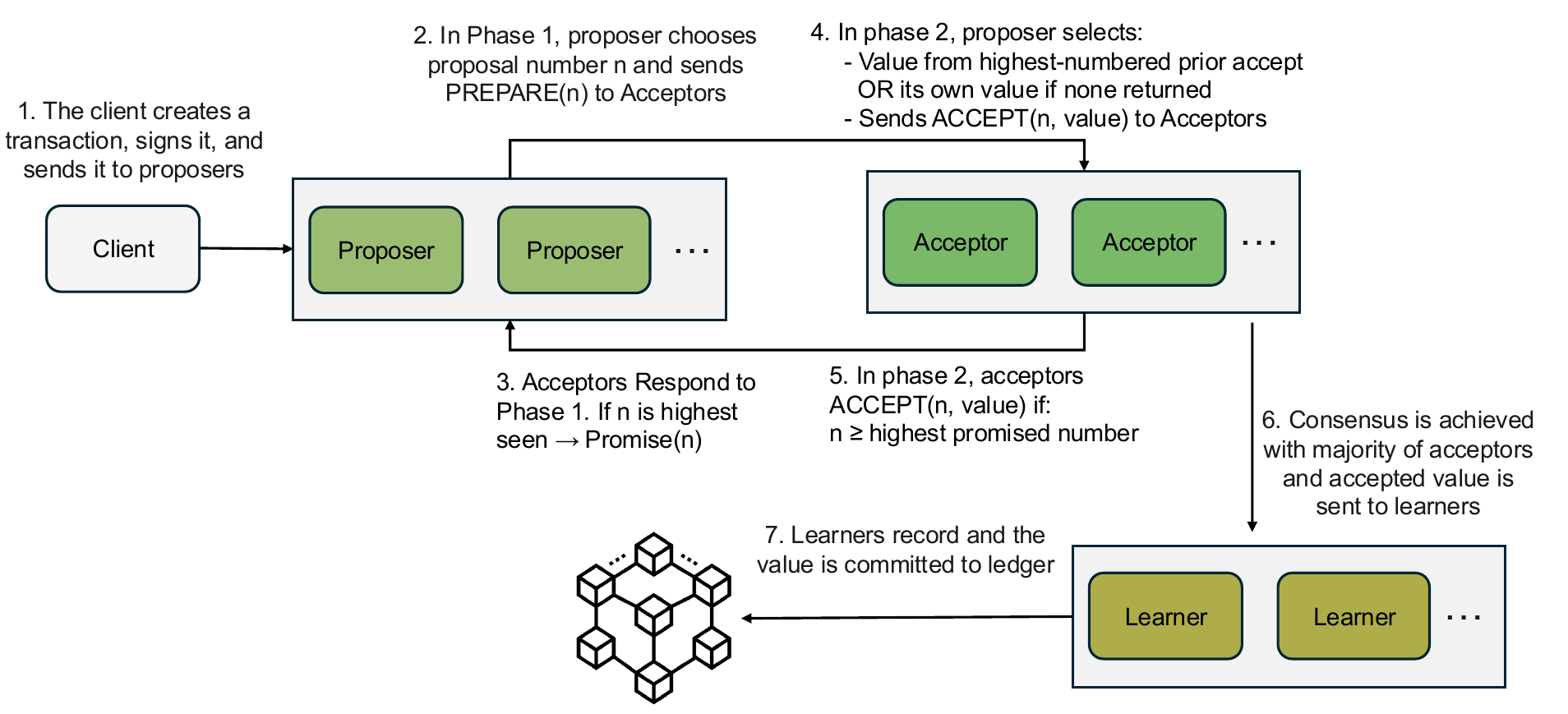}
  \caption{  Consensus workflow on a transaction in the PAXOS protocol}
   \label{fig:paxos-workflow}
\end{figure*}

Paxos is a foundational protocol in distributed systems for achieving consensus on a single value across a network of unreliable processors \cite{garcia2018paxos,tomic2021review1,Official2021-jo}. Its operation, depicted in Figure \ref{fig:paxos-workflow}, involves three roles:
\begin{itemize}
    \item Proposers: Receive client requests and try to persuade acceptors to accept their proposed value.
    \item Acceptors: Form a quorum to accept a proposed value. A proposal is chosen when accepted by a majority of acceptors.
    \item Learners: Learn the chosen value and inform other nodes (e.g., peers in a blockchain).
\end{itemize}

The Paxos consensus mechanism operates through a structured proposal protocol to achieve agreement on a single value across a distributed system. The process begins when a designated proposer node, acting as a leader, receives a client request and formulates a proposal comprising a unique, sequentially-increasing counter and a candidate value. This counter is critical as it allows acceptor nodes to identify and reject outdated or duplicate proposals, ensuring they only consider the most recent request. In the first phase, acceptors evaluate the proposal's counter against their stored history; should it meet or exceed known values, they respond with a promise not to accept older proposals. If a majority of acceptors approve, the proposer broadcasts an accept request. Consensus is achieved once a quorum of acceptors agrees on the value, after which they notify learner nodes, which are responsible for recording the finalized value uniformly across all local ledgers, thus preserving state consistency throughout the network.
While Paxos is renowned for its robustness, it is also complex and difficult to implement. Modern blockchain implementations often use variants of Paxos optimized for logging (e.g., Multi-Paxos).

\section{Comparative Analysis of Consensus Protocols}

This section provides a comprehensive comparative analysis of the consensus protocols discussed in this paper. We evaluate each protocol against a unified set of technical and operational criteria to highlight its fundamental trade-offs. The objective is to provide researchers and practitioners with a clear framework for selecting the appropriate consensus mechanism based on specific security, performance, and governance requirements.


Table \ref{tab:consensus_comparison} provides a functional comparison of the three dominant consensus protocols in public blockchains, highlighting the evolution from the foundational PoW to its more efficient successors. In terms of their approach to Sybil resistance, PoW relies on prohibitively expensive computational work, resulting in high energy consumption but proven, decentralized security. PoS security model is an economic one, where validators' financial stake replaces computational power, drastically improving energy efficiency while maintaining a high degree of decentralization. DPoS further optimizes for performance by introducing a representative democracy, where token holders elect a small set of delegates to validate transactions. This shift enables near-instant finality and high throughput but at the cost of increased centralization, as governance power becomes concentrated in the hands of voters and delegates. The table illustrates the fundamental trade-offs in public blockchain design. PoW maximizes security and decentralization at the expense of scalability, PoS strikes a balance between these axes, and DPoS explicitly prioritizes performance and efficiency, accepting a more semi-centralized governance model as a necessary compromise.

\begin{table*}[htbp]
\centering
\caption{Operational Comparison of Consensus Protocols in Public Blockchains}
\label{tab:consensus_comparison}
\begin{tabularx}{\textwidth}{|l|X|X|X|}
\hline
\textbf{Feature} & \textbf{Proof of Work (PoW)} & \textbf{Proof of Stake (PoS)} & \textbf{Delegated Proof of Stake (DPoS)} \\
\hline
Core Mechanism & Computational competition (hashing) & Economic stake (coin age, randomized selection) & Delegated voting and approval \\
\hline
Primary Incentive & Block reward + transaction fees & Block reward + transaction fees & Block reward + transaction fees \\
\hline
Sybil Resistance & Hardware/Energy cost & Financial stake (coins at risk) & Reputation and voter approval \\
\hline
Energy Consumption & Very High & Low & Very Low \\
\hline
Transaction Finality & Probabilistic (after k-blocks) & Probabilistic \& eventually Absolute (with finality gadgets) & Near-instant \\
\hline
Required Approvals & No fixed percentage; security depends on majority hashpower and number of block confirmations (e.g., 6 in Bitcoin) & At least two-thirds ($\approx$66\%) of validators by stake must attest for finality & At least two-thirds + 1 of delegates ($\approx$70\%) \\
\hline
Governance Model & Fully decentralized, off-chain & Decentralized, off-chain or on-chain & Semi-centralized (token holders elect a small number of block producers) \\
\hline
Example Networks & Bitcoin, Litecoin & Ethereum 2.0, Cardano & EOS, TRON \\
\hline
\end{tabularx}
\end{table*}

Table \ref{tab:private_consensus_comparison} provides a functional comparison of the consensus protocols utilized in private blockchain environments. A primary difference is in the fundamental security model. While PBFT and YAC are designed for Byzantine environments where nodes may act maliciously, Raft, Kafka, Paxos, and PoET only tolerate Crash faults and assume that nodes fail but do not betray the protocol. This directly impacts performance, as seen in the message complexity. PBFT and Paxos require O(n²) communication rounds to achieve consensus in adversarial settings, whereas the CFT protocols (Raft, Kafka, PoET, YAC) achieve linear O(n) complexity, enabling higher throughput and scalability. 
In terms of node role, PBFT employs a straightforward primary-backup structure, while Raft uses a dynamic election system with Leader, Follower, and Candidate states. Kafka's architecture is the most specialized, decoupling the ordering service (Leader/Follower Brokers) from the coordination metadata. PoET and YAC utilize more homogeneous validator nodes, with PoET relying on SGX for leader selection. Paxos features abstract roles including Proposers, Acceptors, and Learners. 
The mechanisms for leader selection also vary, ranging from the view-change rotation in PBFT to democratic, timeout-based elections in Raft and the external coordination via ZooKeeper in Kafka, to the hardware-based random lottery in PoET. 
All these protocols achieve instant, absolute finality, a stark contrast to the probabilistic finality of protocols in public blockchains.
The choice of these protocols by major frameworks demonstrates how their inherent characteristics align with specific application requirements for security, speed, and operational overhead.

\begin{table*}[htbp]
\centering
\caption{Operational Comparison of Consensus Protocols in Private Blockchains}
\label{tab:private_consensus_comparison}
\begin{tabularx}{\textwidth}{|l|X|X|X|X|X|X|}
\hline
\textbf{Feature} & \textbf{PBFT} & \textbf{Raft} & \textbf{Kafka} & \textbf{PoET} & \textbf{YAC} & \textbf{Paxos} \\
\hline
Fault Tolerance & BFT & CFT & CFT & CFT - Trusted Hardware &  BFT &  CFT \\
\hline
Node Roles & Primary, Backup & Leader, Follower, Candidate & Leader Broker, Follower Broker, ZooKeeper & Validators with SGX & Validator nodes, Ordering nodes & Proposers, Acceptors, Learners \\
\hline
Message Complexity & O(n²) & O(n) & O(n) & O(n) & O(n) & O(n²) \\
\hline
Finality & Instant, Absolute & Instant, Absolute & Instant, Absolute & Instant, Absolute & Instant, Absolute & Instant, Absolute \\
\hline
Required Approvals & 2f + 1 (out of 3f+1) & Majority of nodes & Majority of replicas & Majority of active validators & 2f + 1 (out of 3f+1) & Majority of acceptors \\
\hline
Leader Selection & Rotation (view change) & Election (timeout-based) & Appointed (via ZooKeeper) & Random Lottery (via SGX) & Voting-based / round-robin & Election among proposers \\
\hline
Example Frameworks & Hyperledger Fabric, Stellar & Hyperledger Sawtooth, Quorum & Hyperledger Fabric (v1.4) & Hyperledger Sawtooth & Hyperledger Iroha & Many distributed systems (Google Chubby, etc.) \\
\hline
\end{tabularx}
\end{table*}

Table \ref{tab:consensus_protocols} provides a comparative analysis of consensus protocols by quantifying their performance across scalability, latency, energy efficiency and decentralization, and shows a consistent trade-off. Public blockchain protocols like PoW achieve medium-to-high decentralization but at the cost of scalability and latency, with PoW being exceptionally energy-inefficient. Its successors, PoS and DPoS, improve energy efficiency and throughput while maintaining reasonable decentralization, though DPoS accepts a more medium, representative model. In contrast, private blockchain protocols (PBFT, Raft, Kafka) overwhelmingly prioritize performance, achieving very low latency ($<$1 second) and high throughput (thousands to tens of thousands of TPS) with excellent energy efficiency, but they do so by operating in a low-decentralization, consortium-based trust model. YAC occupies a middle ground with moderate scalability and latency, reflecting its design for permissioned voting systems, while PoET's performance is contingent on the trusted hardware (SGX) it relies upon. Overall, the table demonstrates that no single protocol excels in all four categories; the choice involves a strategic compromise, with public protocols prioritizing trust minimization and private protocols optimizing for enterprise-grade speed and efficiency.

\begin{table*}[htbp]
\centering
\caption{Comparison of Scalability, Latency, Energy, and Decentralization}
\label{tab:consensus_protocols}
\begin{tabularx}{\textwidth}{|l|X|X|X|X|}
\hline
\textbf{Protocol} & \textbf{Scalability (TPS)} & \textbf{Latency (Finality Time)} & \textbf{Energy Efficiency} & \textbf{Decentralization} \\
\hline
PoW   & Low (3--15) & High (10--60 min, chain dependent) & Very Poor & Medium--High (affected by mining pools) \\
\hline
PoS   & Low--Medium (10s--1000s) (implementation dependent) & Seconds--Minutes & Excellent & Medium--High (staking centralization risk) \\
\hline
DPoS  & Medium (100s--1000s) & Low (1--3 sec) & Excellent & Medium \\
\hline
PBFT  & High (10K+) & Very Low ($<1$ sec) & Excellent & Low (Consortium) \\
\hline
Raft  & High (10K--50K) & Very Low ($<1$ sec) & Excellent & Low (Consortium) \\
\hline
Kafka & Very High (100K+) & Very Low ($<1$ sec) & Excellent & Very Low (Centralized Ordering) \\
\hline
PoET  & High (10K+) (depends on implementation) & Very Low ($<1$ sec) & Excellent* (relies on SGX) & Low (Consortium) \\
\hline
YAC   & Medium (10K+) & Low ($<3$ sec) & Excellent & Medium (permissioned voting) \\
\hline
Paxos & Medium (100s--1000s) & Very Low ($<1$ sec) & Excellent & Low--Medium (cluster-based) \\
\hline
\end{tabularx}
\end{table*}

\section{Discussion}
This research has provided a systematic examination of consensus protocols, the fundamental components that enable blockchain networks. Our comparative analysis, spanning both public and private blockchains, reveals that the selection of a consensus protocol is not a search for a universally superior solution, but an alignment of a protocol's inherent trade-offs with the specific requirements of a target application. This discussion groups our findings into four themes.

\subsection{The Decentralization, Security, and Performance Trilemma}
Our analysis shows that the blockchain trilemma is real. This means there is always a trade-off between decentralization, security, and performance. None of the protocols we looked at managed to do well in all three at the same time.

Public blockchains, such as those using PoW, explicitly prioritize decentralization and security at the expense of P
performance. PoW's Sybil resistance through physical computation results in robust, permissionless security, but incurs very high energy consumption and very low throughput. PoS and DPoS represent efforts to recalibrate this balance. By replacing physical work with economic stake and delegated governance, they achieve dramatic gains in energy efficiency and transaction speed. However, this comes at a cost: PoS introduces risks of wealth-based centralization, while DPoS explicitly accepts a semi-centralized model, concentrating power in a small set of validators to achieve its high performance.

Conversely, private blockchains universally sacrifice decentralization for high performance and control. Protocols like Kafka and Raft achieve high throughput with instant finality and minimal energy use precisely because they operate within a permissioned consortium of known entities. This centralized trust model eliminates the need for the costly Sybil resistance mechanisms that throttle public networks. The "Low" or "Very Low" decentralization noted in our analysis is the direct price paid for the performance and efficiency these protocols deliver.

\subsection{The Evolution of Trust Models}
At the heart of the trilemma is a key change in the trust anchor.
Public blockchains build trust using cryptography and economic rules. PoW relies on the effort of work done, PoS on the financial stake locked by validators, and DPoS on the votes of token holders. This model enables permissionless participation and censorship resistance, but it also brings problems like high energy use and complex security issues.
In contrast, private protocols anchor trust in legal and reputational authority. The trust is not in the protocol alone but in the pre-established legal identities of the consortium members, enforced by a CA. It allows private blockchains to employ very efficient CFT protocols like Raft, which can't be used in a trustless environment. It also enables the use of BFT protocols like PBFT without performance degradation, as the node count is managed and identities are known. This reliance on off-chain governance and legal recourse is the primary reason for the performance advantages of private blockchains.

\begin{table*}[t]
    \centering
    \caption{Recommendations for Consensus Protocol Deployment Across Key Industrial Sectors.}
    \label{tab:protocol_recommendations}
    \begin{tabular}{|p{3.5cm}|p{4.5cm}|p{2.5cm}|p{6.0cm}|} 
        \hline
        \textbf{Application Domain} & \textbf{Core Operational Demands} & \textbf{Recommended Protocol} & \textbf{Justification for Selection} \\
        \hline
        \hline
        General Financial Ledgers (Public Cryptocurrencies) & Maximal security posture, maximal de-
centralization, resistance to censorship.  & PoW and PoS & PoW establishes trustless operation via energy expenditure, an essential enabler of immutability. PoS in effect
balances security versus the increased scalability required
in contemporary financial ledgers. \\
        \hline
        Supply Chain \& Logistics & High throughput, quasi-instant
finality, known participants. & PBFT or Raft & Designed for high throughput in permissioned environ-
ments. PBFT provides BFT assurances for small consortia,
whereas Raft is naturally easier and faster for internal enterprise chains. \\
        \hline
        Healthcare Record Management & Severe access control, confidentiality
of information (regulatory compliance),
auditability integrity. & PBFT or Raft & PBFT is particularly suited to known nodes (e.g., medical consor-
tia), providing BFT and immediate finality for confidential records.
Raft is used for simplicity in tightly
controlled governance patterns. \\
        \hline
        Inter-Bank \& Internal Settlements & Ultra speed of processing, instant trans-
action finality, minimal exposure of
nodes. & PBFT or Raft & PBFT achieves the necessary rapid finality for small
banking consortia. Raft can be extremely efficient
in private internal banking networks where ab-
solute Byzantine Fault Tolerance is not a spe-
cific requirement. \\
        \hline
        IoT Data \& Micro-transactions & Huge scalability, negligible energy
footprint, close-to-zero transaction
fees. & PoET & PoET utilizes trusted execution environments (such as
SGX) to achieve fair, rapid, and low-energy consensus,
and hence is uniquely suited for resource-restricted IoT devices. \\
        \hline
    \end{tabular}
\end{table*}

\subsection{Fault Tolerance: A Critical Design Choice}
Our analysis underscores a critical distinction between BFT and CFT that must guide protocol selection.
BFT protocols are designed for adversarial environments and can withstand malicious behavior from a subset of nodes. PBFT provides robust security but suffers from high communication overhead (O(n²)), limiting its scalability. Its use is essential in multi-organizational consortia where participants may have competing interests and cannot be fully trusted.

CFT protocols are designed for environments where nodes are trusted but may fail or crash. They offer superior performance and simpler implementation (O(n) complexity) but provide no defense against malicious actors. E.g., a single malicious leader in Raft can compromise the entire network. These protocols are suitable for internal enterprise systems or highly trusted partnerships where the primary risk is hardware failure, not sabotage.

The erroneous application of a CFT protocol in a potentially adversarial context represents a severe security risk, while the use of a complex BFT protocol in a fully trusted setting incurs unnecessary performance penalties.

\subsection{Recommendations of Consensus Protocols for Various Application Domains}
The analysis of consensus protocols confirms that no single mechanism provides a universal solution. The optimal choice depends heavily on whether the blockchain network is public or permissioned, as well as the operational requirements of the specific application domain. Public financial systems, such as cryptocurrencies, typically rely on PoW or PoS because these protocols offer strong decentralization and robust Sybil resistance suited to open participation. In contrast, private and consortium blockchain deployments where performance, controlled membership, and regulatory constraints are central, require different consensus approaches.

In the Supply Chain Management domain, where participants are identifiable and network size is limited, high-performance protocols such as PBFT or Raft are commonly preferred. These protocols provide the low latency and high throughput necessary to track commodity flows without introducing operational bottlenecks. Similarly, in Healthcare, where access control, auditability, and data confidentiality are critical, permissioned BFT-style protocols (e.g., modern variants of PBFT) or PoA are favored to support controlled access, auditability, and confidentiality while facilitating secure data exchange among trusted entities.

IoT networks may use lightweight, specialized consensus protocols (e.g., hierarchical Raft variants) to accommodate device constraints; and enterprise identity, trade-finance, or inter-organization workflows frequently adopt permissioned BFT or authority-based consensus to support governance, trust, and efficient coordination.

Because classical BFT protocols scale poorly as the number of participants grows (communication overhead typically grows ~O(n²)), they remain most suitable for small to medium-sized consortia typically involving fewer than 50 nodes.

Overall, the proper selection of a consensus protocol should always reflect the network type, the scale of participation, and the priorities of the application domain such as throughput, latency, confidentiality, regulatory compliance, decentralization, and trust assumptions.

\section{Future Research Directions}
The landscape of blockchain consensus protocols, though well-established in theory, continues to offer opportunities for future research aimed at addressing key limitations and enabling next-generation applications. Several major challenges define the path forward.

The foremost issue is the Decentralization-Security–Performance Trilemma, which remains the central barrier to achieving optimal blockchain performance. Future work should explore new architectures such as advanced sharding and layered protocol designs, that can improve transaction throughput and network scalability without weakening Byzantine fault tolerance or the decentralized trust model that underpins public blockchains \cite{khan2021systematic,zhou2020solutions}.

Another pressing need is to bridge the gap between theoretical design and real-world implementation. This requires comprehensive empirical testing of consensus protocols in realistic, adversarial conditions, beyond simulations, to identify weaknesses and operational challenges in large-scale deployments \cite{benhamouda2023analyzing,garg2023blockchain}. At the same time, the evolution of on-chain governance and decentralization mechanisms is essential. As PoS and DPoS models mature, governance frameworks must prevent dominance, promote fair participation, and ensure transparent, democratic decision-making \cite{li2020comparison}.

Future research should also focus on improving interoperability between consensus mechanisms and other technologies.  Many existing cross-chain bridges are semi-trusted or have vulnerabilities.
Lightweight protocols are needed for resource-limited IoT systems, alongside secure, trust-minimized solutions for cross-chain communication. However, this is challenging due to power, connectivity, and latency constraints. \cite{hrouga2022potentials,taherdoost2022blockchain,inbaraj2020need}. 
As privacy becomes increasingly important, integration of privacy-preserving techniques including zero-knowledge proofs and secure multi-party computation into consensus layers should be investigated. It remains a complex open problem requiring efficient cryptographic constructions \cite{andola2021anonymity,averin2020review}.

Finally, security and threat mitigation remain continuous priorities. Researchers must proactively identify new attack vectors and design formally verified defenses to strengthen the resilience of decentralized systems \cite{li2020survey,chen2020survey,amores2020security}.

In summary, the future of blockchain consensus depends on an interdisciplinary approach that combines cryptography, economic incentives, and scalable system design to build more secure, efficient, and equitable decentralized infrastructures.

\section*{\textbf{\textbf{Conclusion } } }
This study examined the fundamental principles of public and private blockchains and analyzed major consensus protocols used in each. Transaction workflows were illustrated to highlight protocol operations, followed by a comparative evaluation of their performance, strengths, and weaknesses. Finally, key research challenges and open issues in the field were identified, offering valuable directions for future investigations and potential advancements in blockchain consensus design.
In conclusion, the landscape of consensus protocols is not a hierarchy but a diverse ecosystem of specialized solutions. PoW remains the foundation for maximizing decentralization in value-store applications, while PoS and its variants power the next generation of public smart contract platforms. In the private sphere, the choice between robust security and elegant efficiency hinges entirely on the trust assumptions of the business consortium. There is no "best" protocol, only the "most appropriate" one for a well-defined set of requirements, a principle that should guide all future research and development in this field.

\bibliographystyle{ieeetr}
\bibliography{citation} 

\end{document}